\documentclass[pre,twocolumn,showpacs,superscriptaddress] {revtex4}
\usepackage{graphics}
\newcommand{\cu}
{\affiliation{Department of Physics, University of Calcutta,
92 Acharya Prafulla Chandra Road, Kolkata 700009, India}}

\begin{document}
  
 \title
{Phase transitions in a two parameter model of opinion dynamics with random kinetic exchanges}

\author	{  Parongama Sen }
\cu
\begin{abstract}

 Recently, a model of opinion formation with kinetic  exchanges has been 
proposed in which a spontaneous symmetry breaking transition was reported [M. Lallouache et al, Phys. Rev. E, {\bf 82} 056112 (2010)].
We generalise the model to incorporate two parameters, $\lambda$, to    represent 
conviction and $\mu$, to represent the influencing ability of individuals. 
A phase boundary given by $\lambda=1-\mu/2$ is  
obtained  separating the symmetric and symmetry broken
phases: 
the  effect of the influencing term  enhances the possibility of  reaching a  consensus in the society. 
The time scale diverges near the phase  boundary in a power law manner.  The  
order parameter and the condensate also  show  power law growth close to
the phase boundary albeit with different exponents. The
exponents in general change along the phase boundary  indicating a 
non-universality. The relaxation times, however, become constant  with  
increasing system size near the phase boundary indicating the absence of any diverging length scale.
Consistently, the fluctuations  remain finite but  show strong dependence
on the trajectory along which it is estimated.


\end{abstract}

\maketitle

Various models for opinion formation  have been proposed in recent times in which a
collective consensus emerges out of individual opinions \cite{Stauffer,rmp,voter,Sznajd,galam}. Opinion can be assumed to be a discrete or continuous variable and it
dynamically evolves according to the interactions between the individuals.  
The  change in the opinion of  an individual  may be effected by binary interactions or 
may be dictated by  a set of other individuals.
 The evolution usually  leads to a steady state characterised either by 
a homogeneous state where people have  similar  
 opinion  or a heterogeneous
behaviour where people have widely different opinions. 
The interactions of the individuals in  opinion dynamics models can be studied in terms of appropriate
tunable parameters and  it is of interest to observe whether such parameters
can drive a phase transition in the system \cite{ph-tr,sb-ps,galam1,galam2,bkcop}.

While several different schemes have been proposed for possible evolution of opinions, a number of models have adopted the
idea of kinetic exchanges in opinion formation \cite{bkcop,cont1-heg,cont2-deff,cont3-fort,cont4-tosc}.
In one such recently introduced model \cite{bkcop}, the opinions of individuals,  continuously varying from 
-1 to +1, were assumed to change after pairwise interactions. 
A parameter representing ``conviction'' dictated the way opinions  were exchanged. Beyond a threshold value of this parameter,
the opinions of the individuals were seen to reach a consensus while below this, all individuals remained 
in a ``neutral'' state with their opinions attaining a  zero value.
Hence the model shows  spontaneous symmetry breaking. Such 
symmetry breaking transitions have also been observed earlier \cite{galam1,galam2},
in  binary opinion models in which the ordered phase is one in which a consensus is reached. This is a symmetry broken phase while in the disordered phase, 
the opinions average out to zero. In \cite{bkcop}, the so called disordered phase is very special in
the sense the opinion of each individual is identically zero making
the average equal to zero trivially. Obviously there is no degeneracy in this
case  and neither any  fluctuation. 
Here, at the  transition point,  evidence of the existence of a  diverging 
timescale was also confirmed.

We introduce in this paper, a generalisation of the above model
where   a second parameter, representing the ``influencing'' capability of 
individuals is incorporated. So  individuals are now 
characterised by  two parameters, one which represents one's belief in one's own opinion (conviction) and the other the ability to influence others.
In \cite{bkcop}, these two qualities were identical. We argue that in general these may be different, e.g.,  a person with a firm belief might not have 
the same degree of influencing  others. We keep
things general allowing both parameters to freely vary from 0 to 1 and investigate the phase transitions in the two
dimensional parameter space. 
Other variations of the basic model have also been considered recently \cite{soumya}.

In the present model, let $x_i (t)$ be the opinion of the $i$th individual at time $t$; then after 
an interaction of the $i$th and $j$th agents, the opinions of the two individuals are changed according to
\begin{eqnarray}
x_i (t+1) = \lambda_i x_i(t) + \epsilon_1 \mu_j x_j(t)\\
x_j (t+1) = \lambda_j x_j (t) + \epsilon_2 \mu_i x_i(t),
\end{eqnarray}
where $\epsilon_1$ and $\epsilon_2$ are independent random variables 
ranging from zero to one. Making $\lambda_i = \mu_i$, one gets back the 
model of \cite{bkcop}.
The opinions of both the individuals are changed at the same time, and the interacting 
individuals are chosen randomly.

In the simplest picture, we keep the two parameters $\lambda$ and $\mu$ independent of the agents, i.e,   
assume a homogeneous population having identical $\lambda$ and $\mu$.

The introduction of the parameter $\mu$ different from $\lambda$ 
lends a different connotation to  the original model studied in \cite{bkcop}.  
One can now interpret the first term containing 
$\lambda$  as a   self interaction term and the 
 term containing $\mu$ representing the influence from others   as an interaction term. 

 A proper  order parameter for this model is $m = |\sum_i x_i|/N$ in analogy
with magnetic system. Hence in the disordered state, $m=0$. In case of \cite{bkcop},  one always ends up with either all zero opinions (in the disordered state) or all positive/negative opinions in the ordered state. 
This is in contrast
with other models of spontaneous symmetry breaking \cite{galam1,galam2} where the disordered
state is a mixture of different opinions. However, a disordered state where
the average opinion is zero {\it {and}} the individuals have identical opinions can only be possible  in models with continuous opinions distributed with both positive
and negative values. 
This so called disordered state with all $x_i =0$ can be regarded as a  special case of 
paramagnetic state in the magnetic language. The difference between this state and a general 
paramagnetic state is that there is no fluctuations here. 

Let us first discuss an extreme limiting case of the present model.
The aim in this kind of models is to look at the dynamical evolution 
starting from a completely random state.
Interestingly, when $\mu=0$ and   
for any $\lambda < 1$, $x_i$ being less than or equal to 1 in magnitude, will
rapidly vanish. However, for $\lambda =1, \mu=0$, opinions will not 
evolve at all. Hence  at this particular point, although $m=0$, the fluctuations
remain nonzero in the thermodynamic limit. Such a situation is not possible to realise in the case 
when $\lambda = \mu$. We will discuss other interesting features of the model
close to this point later in the paper.

In the steady state, we get the condition for nonzero solutions of $\langle x_i \rangle$ as,
\begin{equation}
(1-\lambda )^2 =   \langle\epsilon_1 \epsilon_2 \rangle~\mu^2.
\end{equation}
Since $\epsilon_1 $ and $\epsilon_2$ are independent random variables with mean value equal to 0.5, 
and  as $\lambda, \mu$ cannot exceed 1,  the  above condition reduces to
\begin{equation}
\lambda = 1-\mu/2.
\label{eq-ph-boundary}
\end{equation}

We have carried out simulations for a population of $N = 2^m$ individuals
in general (starting from $N=32$),   the essential results are found to be 
independent  of system size beyond $N=128$.
In the simulations with $N$ individuals  whose opinions are randomly distributed initially, we first investigate the steady state behaviour    and
find that indeed, there is a threshold phenomena as the average opinions shows spontaneous
symmetry breaking above a phase boundary occurring in the $\lambda-\mu$ plane; the phase boundary obtained
numerically matches exactly with (\ref{eq-ph-boundary}). 

For $\lambda=1$, we find that the final state is not only ordered, but is completely polarised in the sense that the opinions of all the individuals are equal 
and exactly $ 1$ (or $-1$)
for all values  of $\mu$ except $\mu=0$ (which we have already discussed).
Thus at $\lambda =1, \mu=0$, we have a sharp discontinuity in $m$. 
 So there is a line $\lambda=1$ in the ordered phase  where the fluctuations vanish completely.  For 
other values of $\lambda  \neq 1$, the nature of the phase is dictated by both $\mu$ and $\lambda$.

The implication of the above observation is quite intricate: it shows that
in case we have a model society where everyone is fully convinced about one's 
opinion, the minimum interaction will be able to make the whole
population  polarised 
completely and perfectly (i.e., with opinions exactly equal to $+1$ or  $-1$).
Since a society without interactions is not conceivable, and that such 
polarisations seldom take place, it has to be concluded that 
$\lambda=1$ is indeed an unrealistic  idealised value. 

\begin{figure}
\rotatebox{0}{\resizebox*{8cm}{!}{\includegraphics{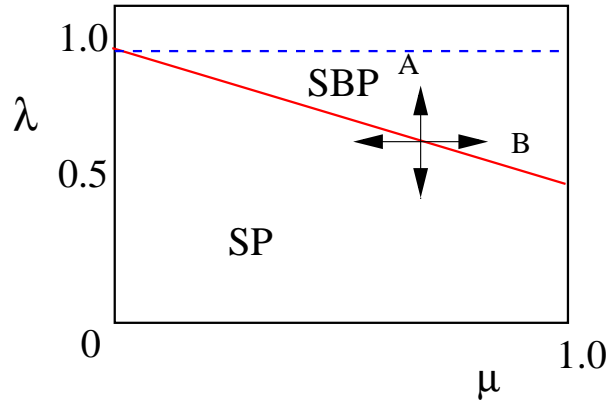}}}
\caption{(Color-online) The phase boundary obtained by numerical simulation coincides exactly with that given in eq (\ref{eq-ph-boundary}).
SP denotes the symmetric phase and SBP the symmetry broken phase.
The paths A and B are possible trajectories along which the different studies can be made. Along the dashed line $\lambda=1$,
the opinions of all the agents are equal and take extreme values in two
possible ways,  
 either $x_i = 1$ or  $x_i = -1$ for all $i$.
 }
\label{ph-dia}
\end{figure}

The phase diagram is plotted in Fig. \ref{ph-dia}. We investigate the nature of the transition at different points on the 
phase boundary. This is done by varying
the parameters close to the transition points ($\lambda_c, \mu_c$) on the phase boundary 
and can be done in several ways  in a two dimensional
plane. We choose two trajectories: path A, where we keep $\mu$ fixed at $\mu_c$ and vary $\lambda$, and 
path B, where $\lambda$ is fixed at $\lambda_c$ and $\mu$ is varied.
In some  special cases, all possible trajectories cannot
be explored, e.g.,  for  $\lambda_c=1, \mu_c = 0$,  the path A does not exist.

Unless explicitly mentioned, the initial state is always taken to be completely
random. The static and dynamical results along both patha A and B have 
been obtained and compared.
To  check the divergence of timescales close to the transition point, 
we find that  the order parameter  given by the average opinion denoted by $m = |\sum_i x_i|/N$ behaves 
conventionally as
\begin{equation}
m(t) \propto  \exp(-t/\tau)
\end{equation}
in the symmetric or disordered phase where $m(t \to \infty) =0$, and 
as
\begin{equation}
m = m_0(1-e^{-t/\tau})
\end{equation}
in the ordered or symmetry broken phase.
In the ordered phase, 
$m_0$, the equilibrium value attained at large times,  is  estimated by averaging $m(t)$ over the last few hundred time steps.

{\it {Results for path A:}}

\begin{figure}
\rotatebox{270}{\resizebox*{6cm}{!}{\includegraphics{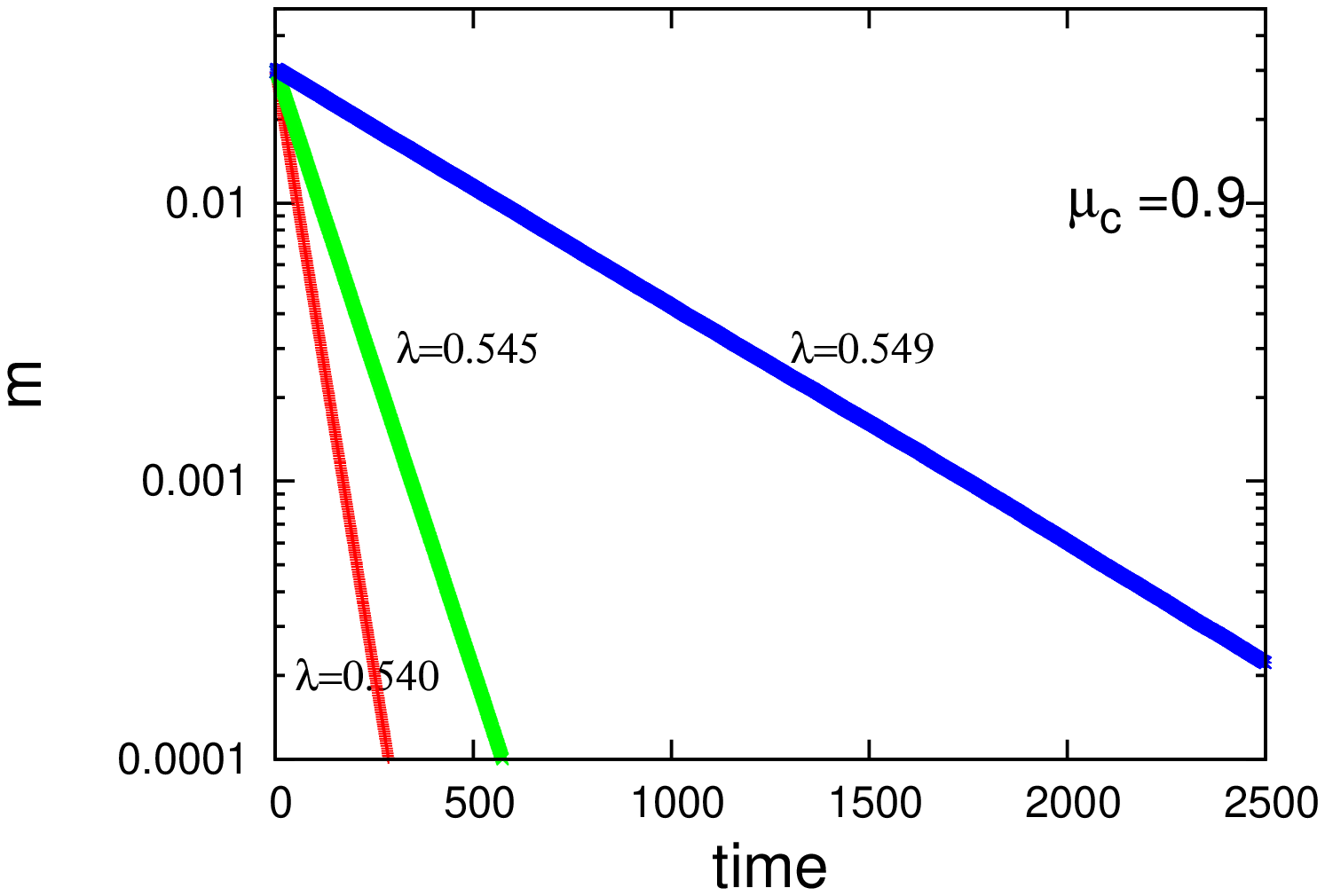}}}
\rotatebox{270}{\resizebox*{6cm}{!}{\includegraphics{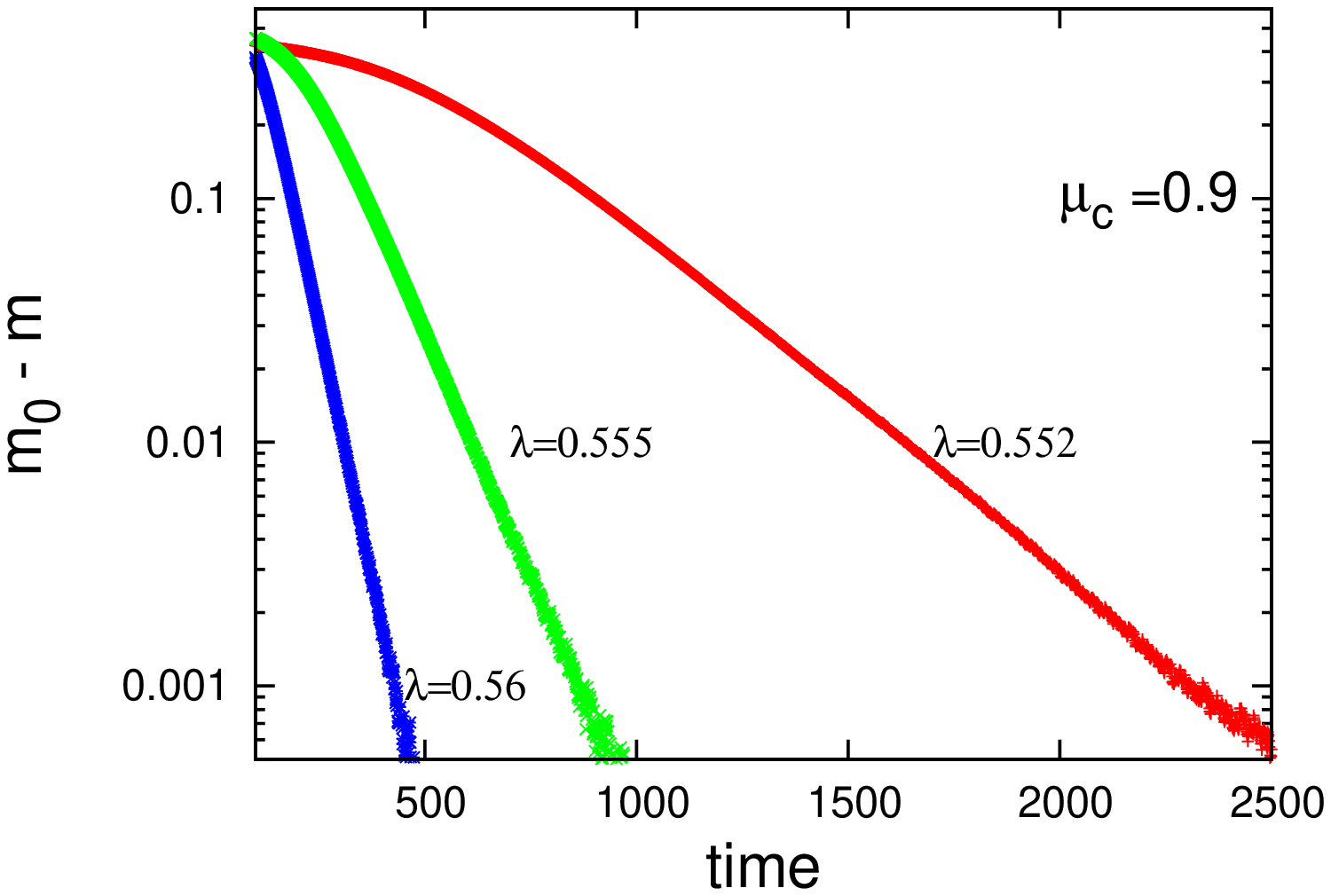}}}
\caption{(Color-online)  The behaviour of the order parameter along path A for $\lambda < \lambda_c$ and 
 $\lambda > \lambda_c$ are shown for a system of $N=256$.}
\label{order}
\end{figure}

\begin{figure}
\rotatebox{0}{\resizebox*{12cm}{!}{\includegraphics{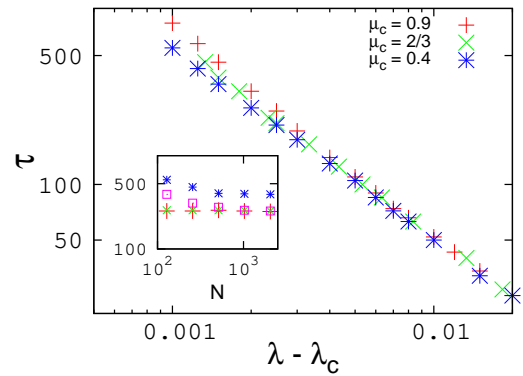}}}{b}
\caption{(Color-online) The timescale variation along path A is shown for three different values of $\mu_c$ for $N=256$ for the order parameter $m$.
Inset shows the values of $\tau$ against the system size $N$ for $\lambda-\lambda_c$ = 0.013 (*) and
$\lambda-\lambda_c$ = 0.02 (other symbols) }
\label{tau-patha}
\end{figure}

In Fig. \ref{order}, we plot   $m(t)$ in the disordered phase and 
$m_0-m(t)$ in the ordered phase to get the exponential decay in both cases. 
Estimating $\tau$ from the slopes of these curves, we show the variation of $\tau$
with $|\lambda -\lambda_c|$ at different locations on the phase boundary 
given by $\lambda_c, \mu_c$ in Fig \ref{tau-patha}. (The locations are indicated by the values of $\mu_c$ only as $\lambda_c$ is related to $\mu_c$ and the
latter is kept constant.) 
 There is a power law variation:
\begin{equation}
\tau \propto (\lambda-\lambda_c)^{-\rho}
\end{equation}
which is true for  both $\lambda < \lambda_c$ and  $\lambda > \lambda_c$ with same values of $\rho$.

 The value of the 
exponent $\rho$  changes very slowly along the phase boundary indicating a non-universal behaviour.
$\rho$  varies systematically as $\mu_c$ is increased; e.g., 
for $\mu_c = 0.4$, $\rho = 1.04 \pm 0.01$, for $\mu_c = 2/3, \rho = 1.10 \pm 0.03$ while for $\mu_c = 0.9$, $\rho = 1.21 \pm 0.01$.

To check the effect of finite sizes on the time scales, we estimate $\tau$ very close to the
phase boundary for different system sizes. Plotted in the inset of Fig. \ref{tau-patha},  the result shows an  interesting
variation. The timescales are larger for smaller system sizes and converge for larger sizes. 
We conclude that the timescales are independent of the system size as it increases.

 We have  studied the behaviour of some other quantities close to the phase transition point also.
The equilibrium value of the order parameter $m$  shows a power law behaviour with  $(\lambda-\lambda_c)$ for $\lambda > \lambda_c$,
\begin{equation}
m \propto (\lambda-\lambda_c)^{\beta}
\end{equation}
where $\beta$ is quite strongly dependent on the point on the phase boundary, e.g., $\beta = 0.079 \pm 0.001$ at
$\mu_c = 0.4$ and $\beta = 0.155 \pm 0.001$ at $\mu_c=0.9$. 
This result again supports the claim that the phase transition is non-universal.

\begin{figure}
\rotatebox{270}{\resizebox*{6cm}{!}{\includegraphics{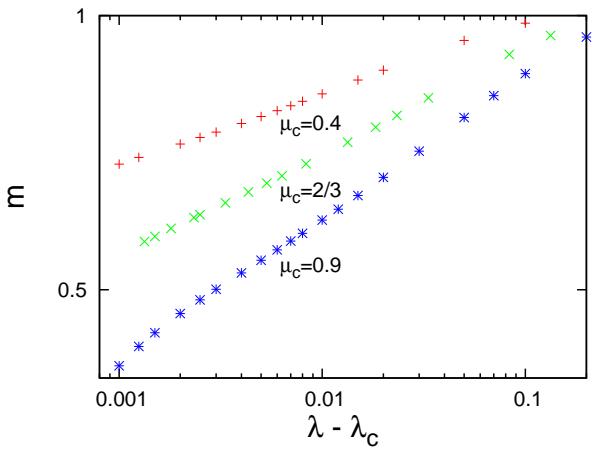}}}
\caption{(Color-online) The equilibrium value of order parameter as a function of $\lambda-\lambda_c$ is shown for different values of $\mu_c$ along path A
for $N=256$. The
exponent is seen to be appreciably dependent on $\mu_c$.}
\end{figure}

\begin{figure}
\rotatebox{270}{\resizebox*{6 cm}{!}{\includegraphics{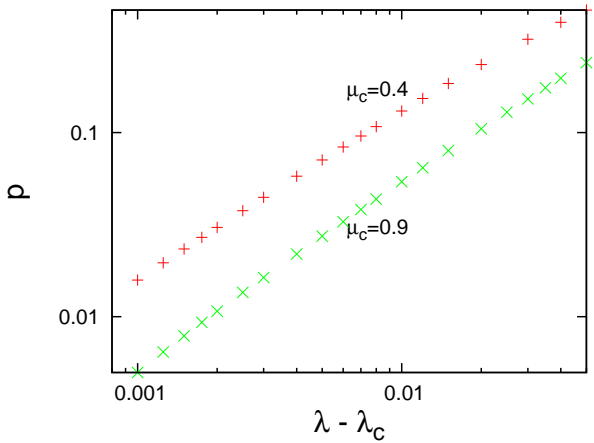}}}
\caption{(Color-online) The equilibrium value of the condensate fraction $p$ 
 as a function of $\lambda-\lambda_c$ is shown for two different values of $\mu_c$ along path A
for $N=256$. The
exponent is seen to be weakly dependent on $\mu_c$.}
\label{mean256.p}
\end{figure}

Another interesting quantity in this type of models is  $p$, the so called 
condensate fraction,
 defined as the fraction of 
individuals having extreme opinions $\pm 1$. We find that it shows a 
behaviour similar to $m$, i.e., it is zero below the critical point and finite above it.  The equilibrium values show scaling with $\lambda -\lambda_c$ for
$\lambda > \lambda_c $ with an associated  exponent $\beta_p \neq \beta$. Once again the values of $\beta_p$ show  nonuniversality but the
nonuniversality is much weaker compared to that found for $\beta$ (see Fig. \ref{mean256.p}). 
For example, $\beta_p \simeq 0.91$ for $\mu_c =0.4$, $\simeq 0.95$ for
$\mu_c =2/3$ (this value agrees with another estimate \cite{soumya}) 
 and $\simeq 1.0$ for $\mu = 0.9$. 

\begin{figure}
\rotatebox{270}{\resizebox*{6 cm}{!}{\includegraphics{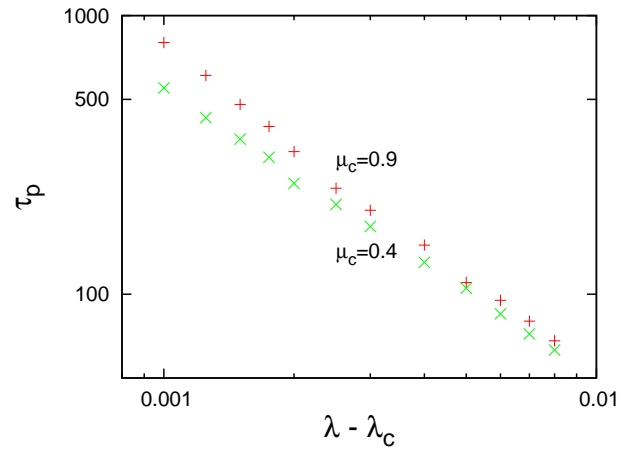}}}
\caption{(Color-online) The timescale variation along path A is shown for two different values of $\lambda_c$
for system size $N=256$ for the condensate fraction $p$.
}
\label{rhop}
\end{figure}

The time periods $\tau_p$ also show  scaling with $\lambda-\lambda_c$ as shown in Fig \ref{rhop}.
The values of the corresponding exponent $\rho_p$
are very close to $\rho$ and again has weak nonuniversality. 
These two exponents appeared to be quite different in \cite{bkcop}.  The values  of these exponents depend very sensitively on the
range of fitting, choice of $\lambda_c, \mu_c$ etc.  We have  taken $\lambda_c$ to 
be  that given by (\ref{eq-ph-boundary}) for a given $\mu_c$ and considered
the range $|\lambda-\lambda_c| \leq 0.01$. We have also checked that
the $\rho$ and $\rho_p$ values become closer as system size is increased.  
An independent estimate   \cite{soumya-pc} for $\rho_p$ for the original model ($\lambda_c=\mu_c =2/3$)  is found to be $\sim 1.1$ which  also agrees with our estimate of $\rho_p = 1.16 \pm 0.03$ at that point.
 The fact that
$\rho$ and $\rho_p$ are alomost equal indicates that there is only one timescale
in the system.
  
One can calculate the fluctuations in the order parameter, 
$\sigma = \sqrt{\sum_i x{_i}^2/N - (\sum_i x_i/N )^2}$,  in the steady state. In the symmetric phase, 
it is zero identically as  $x_i =0$ for all $i$. The results   show that it does not diverge as the phase transition point is
approached in the symmetry broken phase. There does appear a peak in the fluctuation, as in \cite{bkcop}, however, there is no power 
law divergence with $(\lambda-\lambda_c )$. In fact the peak moves away from the transition point for increasing values of
$\mu_c$ and hence the behaviour of the fluctuations is  quite unimportant.

\begin{figure}
\rotatebox{270}{\resizebox*{6 cm}{!}{\includegraphics{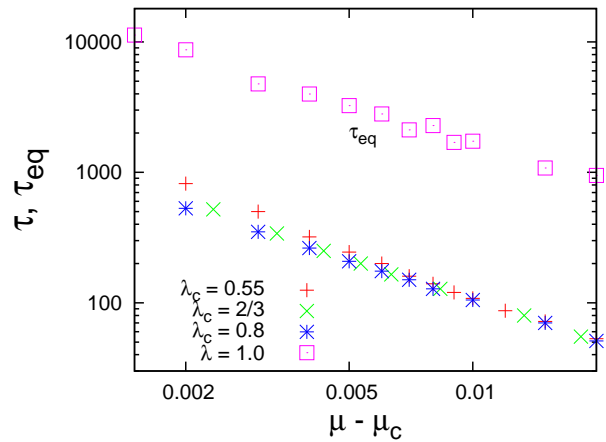}}}
\caption{(Color-online) The timescale ($\tau$) variation along path B is shown for three different values of $\lambda_c$
for system size $N=256$ for $m$; $\tau_{eq}$ is the actual time to equilibriate for $\lambda=1$, for which $\mu_c$ is zero.
}
\label{tau-pathb}
\end{figure}

The facts that the timescales are independent of the system size and the fluctuations do not diverge indicate that 
there is no diverging length scale in the system. That is perhaps because there is no inherent length scales in the system, any 
individual can interact with any other and thus the model is  mean-field like where fluctuations can be ignored.

 {\it {Results along path B}}:

Now $\lambda$ is kept constant, i.e.,    $\lambda = \lambda_c$ and  $\mu$ is varied close to the phase boundary. 
The magnitude of the time scales are about twice compared to those in path B 
along this path, although the value of the exponents are very close.

Along path B, it is possible to study the behaviour close to the special 
point $\lambda=1, \mu=0$. At this  point, there is no order and no 
dynamics. As $\mu$ is switched on, the system relaxes to the steady state where 
the opinion of all individuals are $1$ or $-1$ and it is possible to
calculate the time to reach the steady state. This time, $\tau_{eq}$, is found to be one order of magnitude higher than  $\tau$ or $\tau_p$, and shows a scaling behaviour 
with $\lambda$ with exponent $0.96 \pm 0.03$.  
 The results for $\tau$ and $\tau_{eq}$ 
along path B are shown in Fig \ref{tau-pathb}.

For any quantity $\phi$, one can assume that
\begin{equation}
\phi \propto  (\lambda-\lambda_c)^a \Phi \left[ \frac{\lambda-\lambda_c}{(\mu-\mu_c)^y} \right],
\end{equation}
where $a$ is the exponent obtained in path A and the exponent $b$ from path B is given by $b=ay$. 
The estimate of $y$ from the timescale $\tau$ turns out to be marginally  higher than 1 while that from $m$ is
around 0.9 and is not dependent on the critical values of $\mu_c$ and $\lambda_c$. We conclude that $y$ is 
a universal exponent with a value close to 1.

\begin{figure}
\rotatebox{270}{\resizebox*{6cm}{!}{\includegraphics{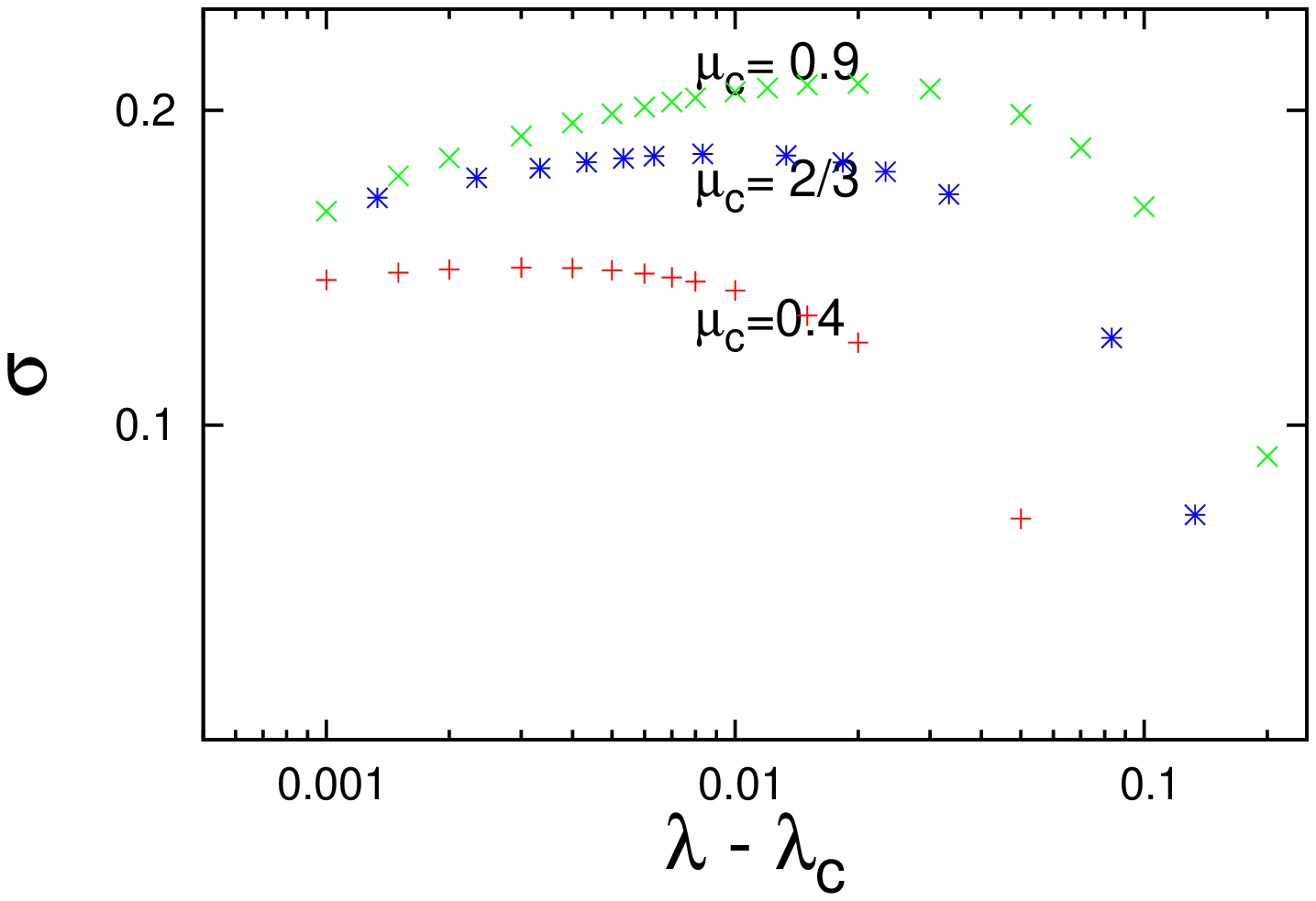}}}
\rotatebox{270}{\resizebox*{6cm}{!}{\includegraphics{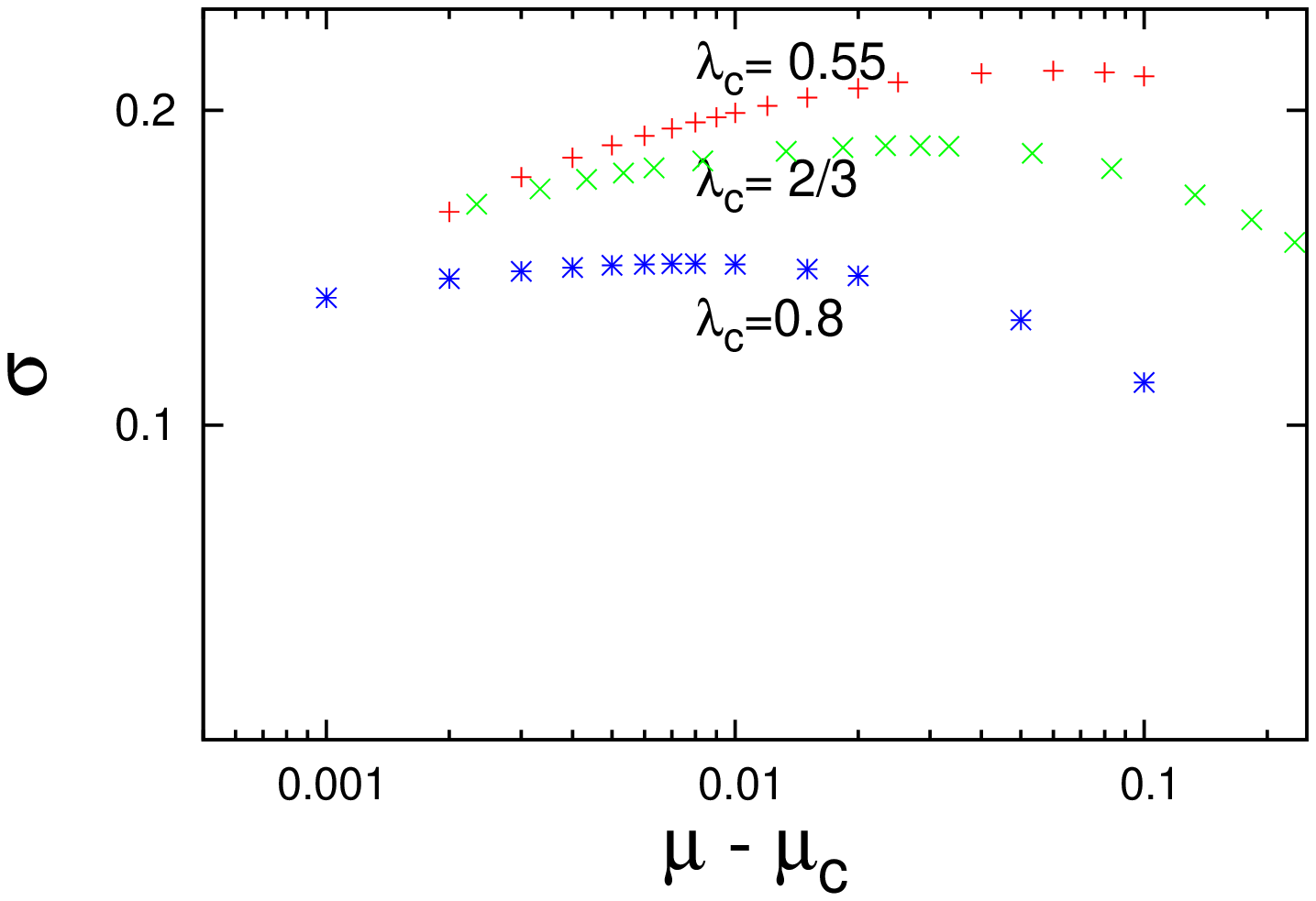}}}
\caption{(Color-online) The equilibrium fluctuations in the order parameter shows no divergence for either path A (top) or B (bottom). The results are
shown for $N=256$.}
\label{fluc}
\end{figure}

The fluctuations, as in path A, once again show a peak which is further away from the transition point and remain finite 
even as $\mu \to 1$, e.g., for $\lambda_c = 0.55$. This is a major difference with path A, where the fluctuations become zero 
far away from the phase boundary at
 $\lambda=1$. 
The fluctuations for both paths are plotted in Fig \ref{fluc}.

The fluctuations in $p$ can also be calculated as it varies from
configuration to configuration.  The results are exactly similar to that of $m$.

So far, we discussed the   dynamics when the initial configuration is 
completely random.
 In case one studies the nonequilibrium relaxation behaviour 
starting from a partially/fully ordered state, a power law decay of the
order parameter is expected exactly at the critical point. This 
exponent, however, is not an independent one as it is given by $\beta/\rho$ (for
$m$) or $\beta_p/\rho_p$ (for $p$).
We have checked that  for both $m$ and $p$   
power law decays with exponents consistent with the already obtained values of $\beta~~ (\beta_p)$ and $\rho ~~(\rho_p)$ can be observed.

In summary, we have studied a model of opinion formation with kinetic exchanges, in which 
two realistic parameters are used. A phase boundary is obtained in the parameter space. 
There are two absorbing phases: the symmetric phase and the extreme boundary line $\lambda=1$.
The model
 is  a generalisation of a recently proposed model with a single parameter representing conviction 
in one's opinion. While conviction is expected to restore one's opinion, the influence of others
may try to change it. Hence these two factors appear in the dynamical evolution with a 
randomness represented by $\epsilon$. The ability to  influence need not be identical to one's conviction. 
Representing the two qualities of 
conviction and influencing as independent parameters, we have shown that 
the effect of the influencing term is like an interaction which enhances the 
possibility of reaching a consensus: cooperative interaction
is reponsible for any order to exist.
The  results show that 
the behaviour of this model with continuous variation  of opinions
is comaparable to binary opinion models only at a special 
point where there is coexistence of many opinions but unlike 
that in \cite{galam1}, it is an absorbing state. In fact  the entire disordered 
phase is an absorbing phase in contrast to the binary opinion models.  

The phase transition along the boundary is shown to be non-universal as the exponents have different values at different points on   the boundary. 
The nonunivrsality is maximum for the order parameter $m$, while for the
other quantities it is weak. The actual values indicate that with larger values of $\mu$, consensus is reached in lesser timescales.  It is still not very clear why the nonuniversal behaviour is there but it perhaps indicates  that the phase boundary is actually  
a line of critical points.

Acknowledgments: The author is grateful to Soumyajyoti Biswas, Anjan Chandra and Arnab Chatterjee for very useful discussions and for sharing their results
prior to publication and to  Abdus Salam ICTP for hospitality where a major part of the work was done.  She also thanks S M Bhattacharjee for some comments. 
Financial support from UPE project is acknowledged.

\end{document}